\def\year{2020}\relax
\documentclass[letterpaper]{article}
\usepackage{aaai20}
\usepackage{times}
\usepackage{helvet}
\usepackage{courier}

\usepackage[hyphens]{url}
\usepackage{graphicx}

\usepackage{tabularx}
\usepackage{amsmath}
\usepackage{subcaption}
\usepackage{booktabs}

\title{Behind the Mask: A Computational Study of Anonymous' Presence on Twitter}
\author{Keenan Jones, Jason R.~C.~Nurse, Shujun Li \\
School of Computing \& Kent Interdisciplinary Research Centre in Cyber Security (KirCCS)\\ 
University of Kent, UK \\
\{ksj5, j.r.c.nurse, s.j.li\}@kent.ac.uk
}
  
\begin{document} 

\maketitle 

\begin{abstract} 
The hacktivist group Anonymous is unusual in its public-facing nature. Unlike other cybercriminal groups, which rely on secrecy and privacy for protection, Anonymous is prevalent on the social media site, Twitter. In this paper we re-examine some key findings reported in previous small-scale qualitative studies of the group using a large-scale computational analysis of Anonymous' presence on Twitter. We specifically refer to reports which reject the group's claims of leaderlessness, and indicate a fracturing of the group after the arrests of prominent members in 2011-2013. In our research, we present the first attempts to use machine learning to identify and analyse the presence of a network of over 20,000 Anonymous accounts spanning from 2008-2019 on the Twitter platform. In turn, this research utilises social network analysis (SNA) and centrality measures to examine the distribution of influence within this large network, identifying the presence of a small number of highly influential accounts. Moreover, we present the first study of tweets from some of the identified key influencer accounts and, through the use of topic modelling, demonstrate a similarity in overarching subjects of discussion between these prominent accounts. These findings provide robust, quantitative evidence to support the claims of smaller-scale, qualitative studies of the Anonymous collective.
\end{abstract}

\section{Introduction}

The hacker/hacktivist collective Anonymous is a group whose nebulous and contradictory ethos provide a source of both bafflement and fascination to those endeavoring to study them. Originating from the /b/ board of the image sharing site 4Chan, in which the board's participants interact anonymously with each other, this sharing of a singular ``Anon'' (a member of Anonymous) identity began to resonate with participants on this site, setting the stage for growth into the group we know today~\cite{Goode2015}. A group famed for their campaigns (dubbed `Ops') targeting many organisations such as The Church of Scientology~\cite{Goode2015}, the security firm HBGary~\cite{Olson2013}, as well as ISIS, and the governments of the United States and Australia~\cite{Goode2015}.

Interestingly, there are a significant number of Twitter accounts claiming some form of affiliation with Anonymous. From these accounts' interactions and posts, one can begin to establish a sense of how the structure and message of Anonymous as a group is presented. Accordingly, in this research we aim to use the findings of our large-scale study of Anonymous Twitter accounts to examine the contentions of smaller-scale, often interview-focused studies of the group. Such studies reject claims by the group to its nebulous, leaderless nature~\cite{Uitermark2017,Olson2013}. Furthermore, they suggest that Anonymous fractured as the result of the arrests of key affiliates~\cite{Goode2015,Olson2013}; a factor which again refutes the argument of a decentralised group structure. 

To achieve this aim, this paper uses computational methods -- specifically machine learning classifiers, social network analysis (SNA), and topic modelling -- to investigate how the findings of qualitative studies of the group, whose results are largely derived from interviews and the examination of secondary sources (i.e., newspaper reports)~\cite{Olson2013,Uitermark2017}, compare to a larger-scale study of Anonymous' actual behaviours on Twitter. Specifically, through our work, this paper:

\begin{itemize}
    \item Identifies the presence of a sizeable network of Anonymous Twitter accounts -- containing more than 20,000 Anons -- using machine learning methods.
    \item Uses SNA and centrality measures to map how influence is distributed across the Anonymous network, confirming the findings of smaller-scale studies (e.g., \cite{Olson2013}) that influence is generally the purview of a small number of members.
    \item Examines how this network has changed over time relative to the arrests of key Anons in the 2011-2013 period~\cite{Uitermark2017}. This longer-term study reveals a network that is seeing a rise in account inactivity, and a decrease in new members.
    \item Compares the overarching tweet content of `key' influencer accounts using topic modelling, finding that each account's tweets follow similar lines of content. Again strengthening the findings of smaller-scale, qualitative studies.
\end{itemize}

From this, our research's large-scale study of Anonymous on Twitter concludes that, contrary to the group's claims, Anonymous displays a far less organisationally flat structure than the group aspires to. Such findings have been suggested by past smaller-scale studies, but as far as we know our work is the first large-scale study to follow a more systematic and computational approach.

It is expected that the insights provided into this group will be of general interest to researchers, cyber security professionals, members of the law enforcement community and the public, by increasing our overall understanding of amorphous hacktivist groups, and their use of social media.  

\section{Background}
\label{Background}

\citeauthor{Uitermark2017}~(\citeyear{Uitermark2017}), in his qualitative analysis of Anonymous' power dynamics, described the group as follows:
\begin{quote}
Anonymous lacks a central authority, has no foundational ideology, does not represent categorically defined groups, does not consistently endorse ideologies, and has no fixed objective.
\end{quote}
This structure allows for the prevalence of multiple conflicting motivations and ideological goals, with the group advocating for nihilism and idealism, libertarianism and socialism, pranks (often referred to as `lulz' a corruption of LOL (laugh out loud), generally used to describe acts that sought humour at the expense of others) and activism, freedom of speech and the suppression of speech~\cite{Olson2013,Goode2015}. As the group matured over time, these targets and goals began to be referred to as `Ops' by members of Anonymous~\cite{Olson2013}. The inception of these Ops generally followed the nebulous structure that the group subscribed to, with the success of an Op directly tying in to its ability to attract enough Anonymous members to join it~\cite{Olson2013}. 

Instead of describing Anonymous using group-based theory, Beraldo used the phrase ``contentious brand'' -- a group defined, and singularly united, by the ``Anonymous signifier''~\cite{Beraldo2017}. This notion of signification is central to Anonymous, from the Guy Fawkes mask, to the headless businessman, to the grandiose ``We are Legion'' style of communication~\cite{Beraldo2017,Olson2013}. And it is the freely available nature of these methods of identification that allow for this movement to be so amorphous. As highlighted by \citeauthor{Nurse2018}~(\citeyear{Nurse2018}), the Anonymous Twitter account @GroupAnon stated:
``\textit{No, this is not the official \#Anonymous account. There is no official account. We have no central leadership. (Other than the FBI/NSA, joke)}''.
And the central reason there can be no official Anonymous account is that its membership is entirely reliant on the utilisation of the Anonymous brand, something freely available to all, rather than something that can be formally accessed by securing an approved membership.

A point of interest however is that Anonymous' position of having no central-authority has often been contradicted in practice. Although both Olson and Uitermark noted Anonymous' claims to a flat leadership structure, they both suggested that Anonymous' reality -- at least on IRC (internet relay chat) -- was far less a flat structure than one with a clear set of leaders~\cite{Olson2013,Uitermark2017}. Uitermark and Olson described the existence of a `\#Command' room on IRC, in which these self appointed leaders -- without the knowledge of other Anons -- would plan the group's Ops. Moreover, a considerable change was noted in the group after the arrest of several members of the \#Command board (also members of the Anonymous splinter group LulzSec~\cite{Olson2013}) in 2012. After this, \citeauthor{Uitermark2017}~(\citeyear{Uitermark2017}) concluded that the group had fragmented considerably, stating that:
\begin{quote}
Anonymous lived on ...\ as a set of symbols and communication channels ...\ appropriated by a range of different groups for a range of different purposes.
\end{quote}
In turn, the group seemed to have lost the coherence present in its early days, leading to a drastic fall in notable operations and exploits~\cite{Goode2015}.

An additional key point of Anonymous, as elucidated by \citeauthor{Nurse2018}~(\citeyear{Nurse2018}), is that it is a group that has a strong ``public-facing nature''. Their work noted the presence of several Twitter profiles controlled by Anonymous affiliates, and the group further confirmed this via its willingness to engage with journalists; be it through IRC or even via interviews~\cite{Olson2013}. By being public facing, Anonymous can easily capture more media attention, and bring new recruits to whichever Op the group, or a splinter of the group, is executing.

\section{Related Work}

Due to the strong influence that high reputation players can wield on a particular network, it is of interest to be able to identify who these key players are. In~\cite{Nouh2015}, the authors identified a social network of activists operating on Facebook, and then deployed a variety of techniques to identify the key players within this network. This research utilised centrality measures such as betweenness and eigenvector centrality to assess the influence of activists within the group. These metrics give a sense of how interconnected the network is, its durability, and which of the nodes in the network wield the most influence. From this, a disparity between the activity of a user and their influence on the network was identified, as well as the presence of sub-communities with stronger ties within the wider network. In doing so, this work lays out a strong method for gaining a good analytical understanding of influence within a given social network.

Social network graphs, combined with PageRank centrality measures, have also been utilised by \citeauthor{Alfifi2019}~(\citeyear{Alfifi2019}) in their study of ISIS on Twitter. This was done via the construction of a network graph representing the connections between ISIS accounts via their retweets, and subsequently utilised PageRank to estimate the overall influence that ISIS held on Twitter. This was then compared to a randomly sampled group to investigate whether ISIS holds, or is able to exert, more influence over the Twitter community than the average Twitter group~\cite{Alfifi2019}. 

These forms of network analysis have also been applied to the study of Anonymous on Twitter, and the manner in which a network, derived from the use of ``\#Anonymous'', evolved over the period from 1 December 2012 to 30 November 2015~\cite{Beraldo2017}. In turn, the work examined the stability of this Anonymous network, finding that stability in general appeared low, with the number of nodes recurring in the following month falling by 47.8\%, and the percentage of connections between nodes surviving the month being, on average, 12.6\%~\cite{Beraldo2017}. 

Moreover, an additional study of Anonymous on Twitter utilised this same method of identifying Anonymous affiliates by their use of ``\#Anonymous'' to examine differences between male and female account holders, and the types of `Ops' they tweeted about~\cite{McGovern2020}. From this, it was found that male Anons generally tweeted about a multitude of different `Ops' -- including those focused on animal rights activism, and the identification of sexual predators -- whilst female Anons focused almost entirely on `Ops' concerning animal rights~\cite{McGovern2020}. 

Topic modelling is a technique that seeks to extract clusters of words within a document that are semantically similar, in turn dividing the document into the word groupings or `topics' that it is comprised of. This technique has seen use within the field of cybercriminal study including~\cite{Tavabi2019,Kigerl2018} amongst others. \citeauthor{Tavabi2019}~(\citeyear{Tavabi2019}) used  latent Dirichlet allocation (LDA) -- a form of topic modelling that allows words to occur in multiple topics -- to help understand the core content of activity across 80 dark web forums. This included a number of topics related to vending, security and gaming. 

Moreover, \citeauthor{Kigerl2018}~(\citeyear{Kigerl2018}) applied a similar technique to dark-web carding forums (dark-web services centred around the sale of stolen bank cards); identifying the key topics present within the comment histories of users, including keywords pertaining to products sold, and customer satisfaction. This technique therefore helps provide a summative view of a large amount of textual information, allowing for large-scale studies of cybercriminal groups on online platforms. 

\section{Our Contributions}

 Anonymous is unusual in its desire for the spotlight, and the manner in which it relies on social media (e.g., Twitter and YouTube) to profess its messages. Therefore, this paper aims to examine the results of the small-scale, generally interview-based and anecdotal studies of the group seen in Section~\ref{Background}, using a study of a large network of Anonymous affiliated accounts on Twitter. To our knowledge this form of large scale analysis of Anonymous Twitter accounts has not been performed before. Moreover, this article focuses on the more permanent Anonymous affiliated accounts, and in particular, how influence is distributed across the network and how the network has changed over time -- especially in response to the 2011-2013 arrests~\cite{Olson2013}. 
 
 Our research moves beyond previous studies of Anonymous on Twitter which relied on the more narrow use of ``\#Anonymous''~\cite{Beraldo2017,McGovern2020} to identify Anonymous affiliates. Albeit arguable, such methods of characterisation seem to ignore the wide range of purposes for which any account may use a particular hashtag. Moreover, this work expands upon the studies of Anonymous' content, by being the first to conduct a systematic study of Anonymous tweets. This therefore moves beyond the methodology of \citeauthor{McGovern2020}~(\citeyear{McGovern2020}) which focused purely on frequencies of `Op' related hashtag usage. Our paper also addresses the lack of studies focused on the group after the arrests between 2011 and 2013 and its alleged fall from the limelight~\cite{Goode2015,Olson2013}. These analyses provide unique empirical insights into one of the most well-known hacker/hacktivists groups that has ever existed.

\section{Methodology}

In order to meet this study's aims, machine learning was used to construct a network graph detailing the connections between a previously unidentified large sampling of Anonymous accounts. This network was then used to examine the manner in which influence presented itself, and whether there were any identifiable key Anonymous influencers. In turn, the latest tweets of selected influencer accounts were collected and LDA topic modelling conducted. The topics identified were then analysed to gain a sense of the cohesiveness in subject matter across influencer accounts, as well as to identify any similarities between these topics and Anonymous' interests before the aforementioned 2011-2013 arrests. We detail our approach below.

\subsection{Data Collection}

To generate the dataset, two-staged snowball sampling was used to recursively iterate through five Anonymous seed accounts. We collected the Anonymous followers and friends of these five accounts, and thereafter the Anonymous followers and friends of each seed accounts' followers and friends. This technique was chosen for its effectiveness at sampling hidden populations on Twitter~\cite{Benigni2017}. The five accounts selected as seeds were chosen as they had been identified by \citeauthor{Nurse2018}~(\citeyear{Nurse2018}) as being notably linked to the Anonymous collective. The account @YourAnonGlobal, the sixth Anonymous account mentioned in Nurse et al.'s paper, was not included in the study as the account is currently suspended.

All friend and follower extractions were made using Twitter's Standard search API. Twitter restricts the sharing of Twitter content (in this case, account names) should the extracted content subsequently be suspended, deleted, or made private. Therefore, this paper has opted to pseudonymise references to specific accounts to ensure this paper remains compliant with Twitter's terms and conditions should any of these events happen to the pseudonymised accounts referenced in this paper~\cite{TwitterDev}. It is worth clarifying that any account that had been deleted, suspended, or made protected prior to when the sampling was conducted will not have been included in the study as Twitter does not make these accounts available to those using their search API.

As the totalled follower and friend numbers for the five seed accounts was approximately 2 million (as of 1 December 2019), it was necessary to automate the identification of Anonymous accounts. Moreover, due to the nebulous nature of the group, a rigorous definition by which to manually label these accounts as Anonymous affiliated does not exist. It was therefore decided that a rudimentary definition would be used to manually label a number of  Anonymous accounts, before leveraging a machine learning model trained on these labelled accounts to identify Anonymous accounts amongst the followers and friends.

\begin{table}[!tbh]
\centering
\begin{tabular}{ c c c c }
 \toprule
 anonymous & an0nym0u5 & anonymou5 & an0nymous \\
 anonym0us & anonym0u5 & an0nymou5 & an0nym0us \\ 
 anony & an0ny & anon & an0n\\ 
 legion & l3gion & legi0n & le3gi0n\\
 leg1on & l3g1on & leg10n & l3g10n\\
 \bottomrule
\end{tabular}
\caption{Anonymous keywords used.}
\label{table:keywords}
\end{table}

To provide the training set for the model, we decided that a `positive' Anonymous account would be defined as one that contained at least one Anonymous-related keyword (see Table~\ref{table:keywords}) in either its username or screen-name, and in its description, as well as having a profile or background (cover) image containing either a Guy Fawkes mask or a floating businessman (both common Anon images~\cite{Olson2013}). This definition, whilst likely overzealous in its specificity, allowed for a set of Anonymous accounts that are relatively, given the topic, uncontroversial in their labelling. The keywords were derived from the Anonymous literature~\cite{Olson2013,Goode2015}, and a manual examination of the five Anonymous seed accounts.

\begin{table*}[ht!]
\centering
\begin{tabular}{p{0.25\textwidth} p{0.25\textwidth} p{0.3\textwidth}} 
 \toprule
 \textbf{Anonymous Features} & \textbf{Profile Features} & \textbf{Content Features} \\ 
 \midrule
 `Anonymous' used & Tweet number & Number of characters\\
 `Anon' used & Follower number & Number of words\\
 `Anony' used & Friend Number & URL in description\\
 `Legion' used & Follower-Friend ratio & Number of uppercase characters\\
 `Ops' used & Favorites Number & Number of lowercase characters\\
 Anon motto in description & Number of Lists & Number of alphabetical characters\\
 `Hacker' terms used & Location Provided & Number of numerical characters\\
 L33t speak used & Is account protected & Number of punctuation characters\\
 Capitalisation within words & Is URL provided & Number of emoji used\\
  & & Number of mentions \\
  & & Number of hashtags \\
  & & Flesch-Kincaid score \\
  & & VADER score \\
 \bottomrule
\end{tabular}
\caption{List of features.}
\label{table:features}
\end{table*}

The labelling process of the five seed accounts' followers and friends was then carried out using a combination of automated and manual techniques. Firstly, each of the followers and friends were automatically filtered based on the presence of Anonymous keywords (Table~\ref{table:keywords}) in either the username or screen-name of each account. After the filtering process approximately 38,000 accounts remained. These were then hand labelled by the authors of the paper, via a manual examination of each account's Twitter page, in accordance with the prescribed definition. This in turn identified 9,337 Anonymous accounts, and 29,351 non-Anonymous accounts.

These two sets of hand-labelled accounts were then used to train a machine learning model to automatically identify Anonymous accounts. SVM (with a sigmoid kernel), decision trees, and random forest (with 100 trees) were tested to identify the best performer. All testing was conducted using five-fold cross validation. For this research, the Scikit-learn implementations of these algorithms were used~\cite{Pedregosa2011}.

In total, 62 features were extracted from each account, including features directly related to Anonymous and hacker culture, as well as features noting the profile and content information of each account. Moreover, two metrics: Flesch-Kincaid reading-age score~\cite{Aslan2018}, and VADER compound sentiment analysis~\cite{vader} were also calculated from each account's description and the results used as additional features. A full list of the features can be found in Table~\ref{table:features}. For clarity, the Anonymous motto -- referenced in Table~\ref{table:features} -- is detailed in~\cite{Olson2013}. Where appropriate, the features in the list were collected for the username, screen-name, and description of each account (so for number of characters, three separate features are recorded). It is worth emphasising that these features were derived only from the account information; no tweets were leveraged in the creation of this classifier. Whilst this likely inhibited the effectiveness of the model, it also allowed us to test the ability of models to classify accounts despite having a limited amount of information to draw upon. This was a crucial test given the tighter restrictions Twitter has recently enacted in terms of tweet collection using its search APIs.

Having extracted features for the positive and negative sets of accounts, each of the three models were tested using 5-fold cross validation to identify the most effective algorithm. Of the three, random forest offered the highest scores both for the overall accuracy, and for precision and recall with a score of 0.94 for each of these metrics (see Table~\ref{table:ML_performances} for a full comparison of performances). It was thus selected as the algorithm of choice.

\begin{table}
\small
\centering
\begin{tabular}{ c c c c }
\toprule
\textbf{Model} & \textbf{Precision} & \textbf{Recall} & \textbf{F1-Score} \\
\midrule
Random forest & \textbf{0.94} & \textbf{0.94} & \textbf{0.94} \\ 
Decision tree & 0.90 & 0.90 & 0.90\\ 
SVM (sigmoid kernel) & 0.66 & 0.74 & 0.67\\
\bottomrule
\end{tabular}
\caption{Performances of the three tested machine learning models.}
\label{table:ML_performances}
\end{table}

Having established its effectiveness, the random forest model was trained on the entirety of the manually-labelled dataset and was then used on the entire seed follower and friend list to identify the presence of any Anonymous accounts that had been discounted using the crude name filtering during the labelling process. The model identified 15,222 Anonymous accounts in total -- including the approximate 8,846 accounts that had been manually labelled and in turn identified by the model in the training phase. The process was then repeated on the follower and friends of these identified Anonymous accounts, yielding a total of 20,506 identified Anonymous accounts across the two-stages of sampling.

\subsection{Building the Anonymous Network and Identifying Influencers}

Social network analysis (SNA) is an analytical technique that investigates ``social structures by utilising graph theory concepts''~\cite{Nouh2015}. This mapping of connections between actors then allows for the use of network centrality measures to identify the nodes which exert the most influence over the network~\cite{Alfifi2019,Beraldo2017}.

Four of the most common metrics for identifying influencer nodes are degree centrality, eigenvector centrality, its variant PageRank, and betweenness centrality~\cite{Borgatti2005,Nouh2015}. These metrics have seen use in the analysis of social media groups; including both Anonymous and Occupy Wall Street on Twitter, and other activists on Facebook~\cite{Borgatti2005,Tremayne2014,Nouh2015}, and allow for the identification of `key' influencer nodes in a given network.

In our work, SNA is used to map the relationships between identified Anonymous accounts and centrality measures used to identify the key influencer accounts in the network. We defined a non-weighted directed graph \(G = (V,E)\) where each node \(u\in V\) represents an Anonymous Twitter account, and each edge \((p,q)\in E\) a follower relationship, where Anonymous account \(p\) is a follower of Anonymous account \(q\) . In order to obtain the most accurate sense of which accounts are most influential; degree centrality, eigenvector centrality, PageRank, and betweenness centrality scores were recorded for each account in the network. The construction of the SNA graph and the centrality calculations were conducted using the NetworkX Python package~\cite{networkx}.

After calculating the scores for each account in these selected centrality measures, the resulting scores were normalised, and each account's average score across the four measures was recorded. As each metric measures influence in a slightly different way, this method attempts to ensure that the accounts that are the most broadly influential are ranked the highest.

\subsection{Extracting Topics From Key Anonymous Tweets}

Latent Dirichlect allocation (LDA) is among the most popular forms of topic modelling~\cite{Tavabi2019}, and is the method used in this paper. LDA works under the assumption that each given document is comprised of a random mixture of latent topics, with each topic being made up of a particular distribution of words~\cite{Tavabi2019}. LDA allows for terms to be members of more than one topic, and the the assignment is probabilistic, with the probabilities of each term across topics summing to 1.0~\cite{Kigerl2018}. LDA works over a set number of iterations, improving and updating the probabilities upon previous iterations of the model until significant improvement is deemed to have halted. The Gensim Python implementation of LDA was used for this research~\cite{gensim}.

The model itself provides one estimate -- the distribution of words over topics, derived from another, user specified, estimate -- the number of topics present in the document~\cite{Kigerl2018}. As the number of topics with which the LDA model attempts to distribute over are user-specified, it was necessary to run the model continuously over a variety of topic numbers to identity the most suitable topic number per Anonymous account. In order to assess the quality of each topic number, the UCI topic coherence was used as a metric to assess the quality of the learned topics for each given topic-number~\cite{Roder2015}.

As this research sought to carry out topic modelling on the tweets of the most influential Anonymous accounts, the top 6 scoring accounts averaged across the aforementioned centrality measures were selected to be the accounts of interest. The top influencer accounts were selected for further analysis as they are the accounts in the best position to spread their messages, via tweeting, to the whole Anonymous network. Thereby making their messages the most pervasive in the network. 

In turn, the most recent 1,500 tweets of the top 6 accounts were collected using Twitter's Premium search API~\cite{TwitterPremium}  from 1 July 2019 to 29 December 2019. Each account's most recent tweets were collected as this study aims to assess the activities and behaviours of these accounts as of their latest interactions on Twitter. Moreover, as the influencer accounts are identified from their current relations with other accounts, it seemed sensible to examine their most current tweets, rather than tweets from a particular time period. In order to comply with Twitter's developer guidelines the selected accounts will be referred to as accounts A to F; with Account A representing the most influential account, Account B the second, and so on.

\begin{figure}[!tbh]
\centering
\includegraphics[width=\columnwidth, height=0.25\textheight]{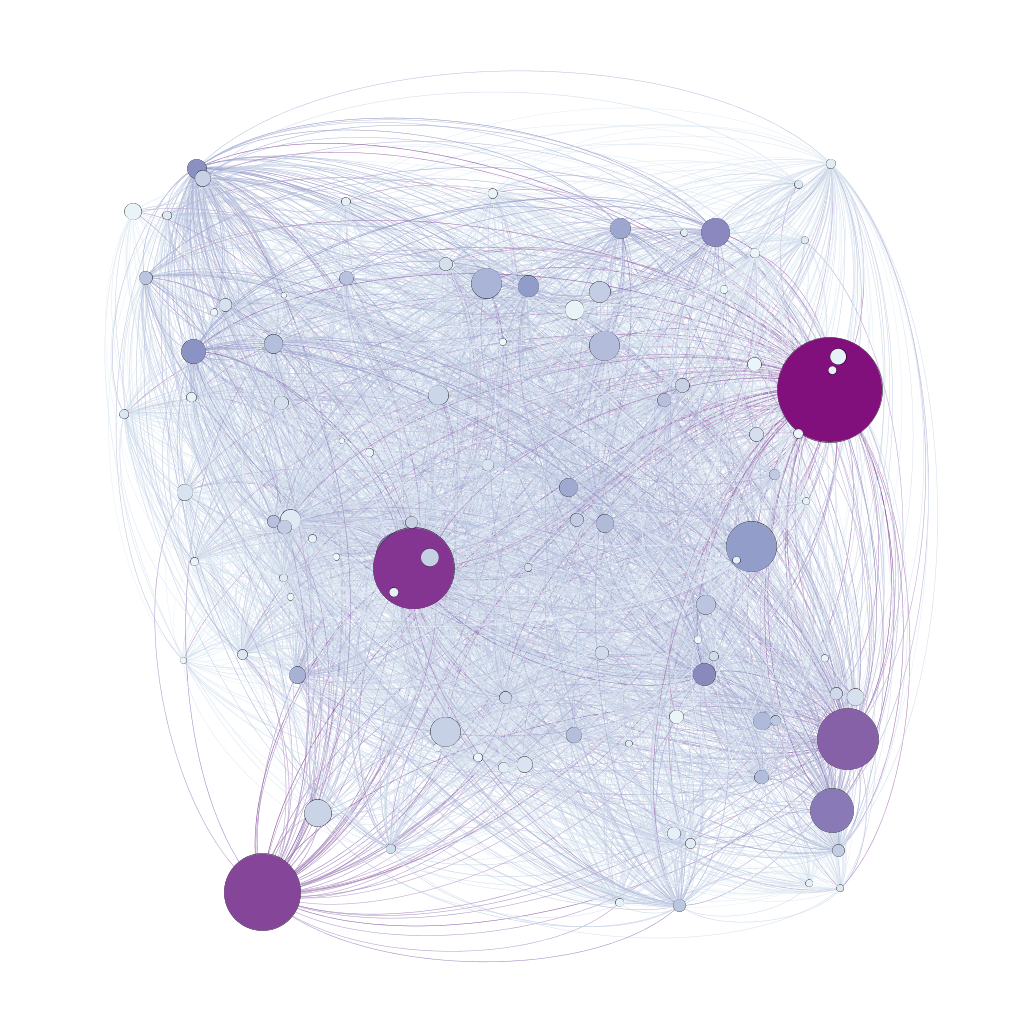}
\caption{Trimmed network of the top 100 Anonymous accounts ranked by eigenvector centrality score. The darkness of node color corresponds to eigenvector score, the node size corresponds to degree centrality score. Note the presence of key nodes that significantly outscore their peers, especially given the reduced network size.}
\label{fig:sna_graph}
\end{figure}

\begin{figure*}[!htp]
\centering
\begin{subfigure}[b]{0.4\textwidth}
    \centering
    \includegraphics[width=\textwidth, height=0.21\textheight]{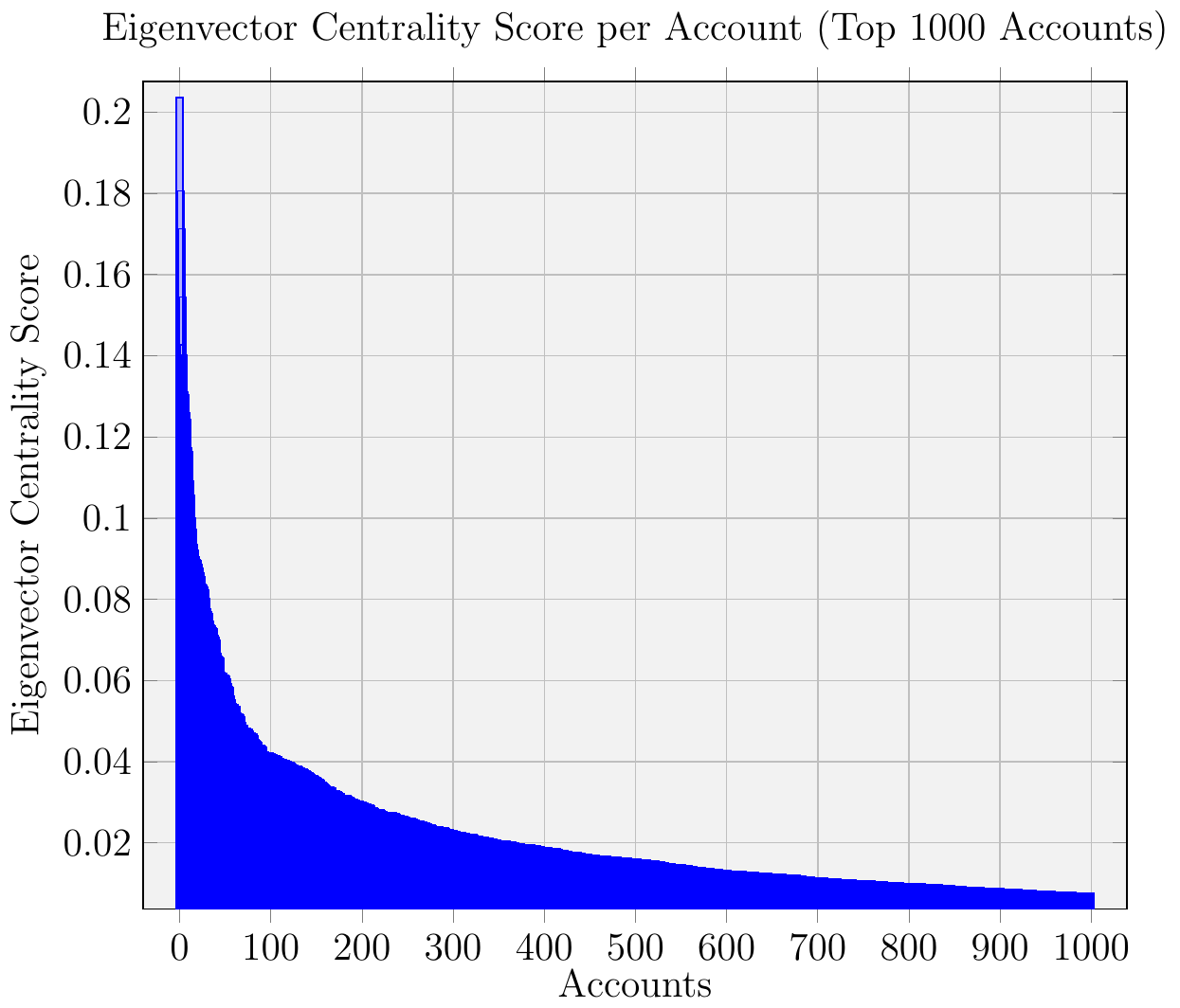}
    \label{fig:eigenCent1000}
\end{subfigure}
\quad
\begin{subfigure}[b]{0.4\textwidth}
     \centering
     \includegraphics[width=\textwidth, height=0.21\textheight]{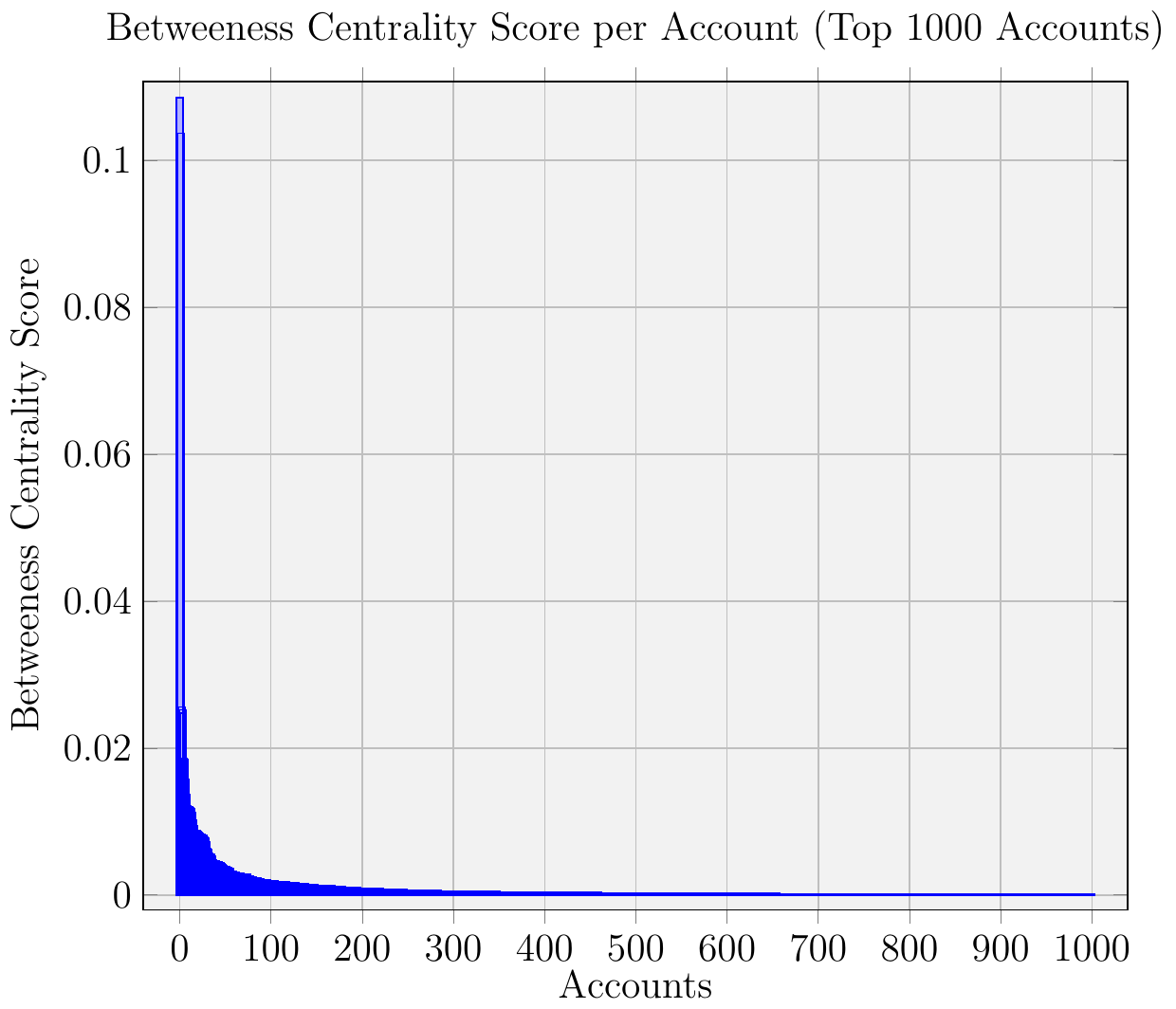}
     \label{fig:betweenCent1000}
\end{subfigure}\\
\begin{subfigure}[b]{0.4\textwidth}
    \centering
    \includegraphics[width=\textwidth, height=0.21\textheight]{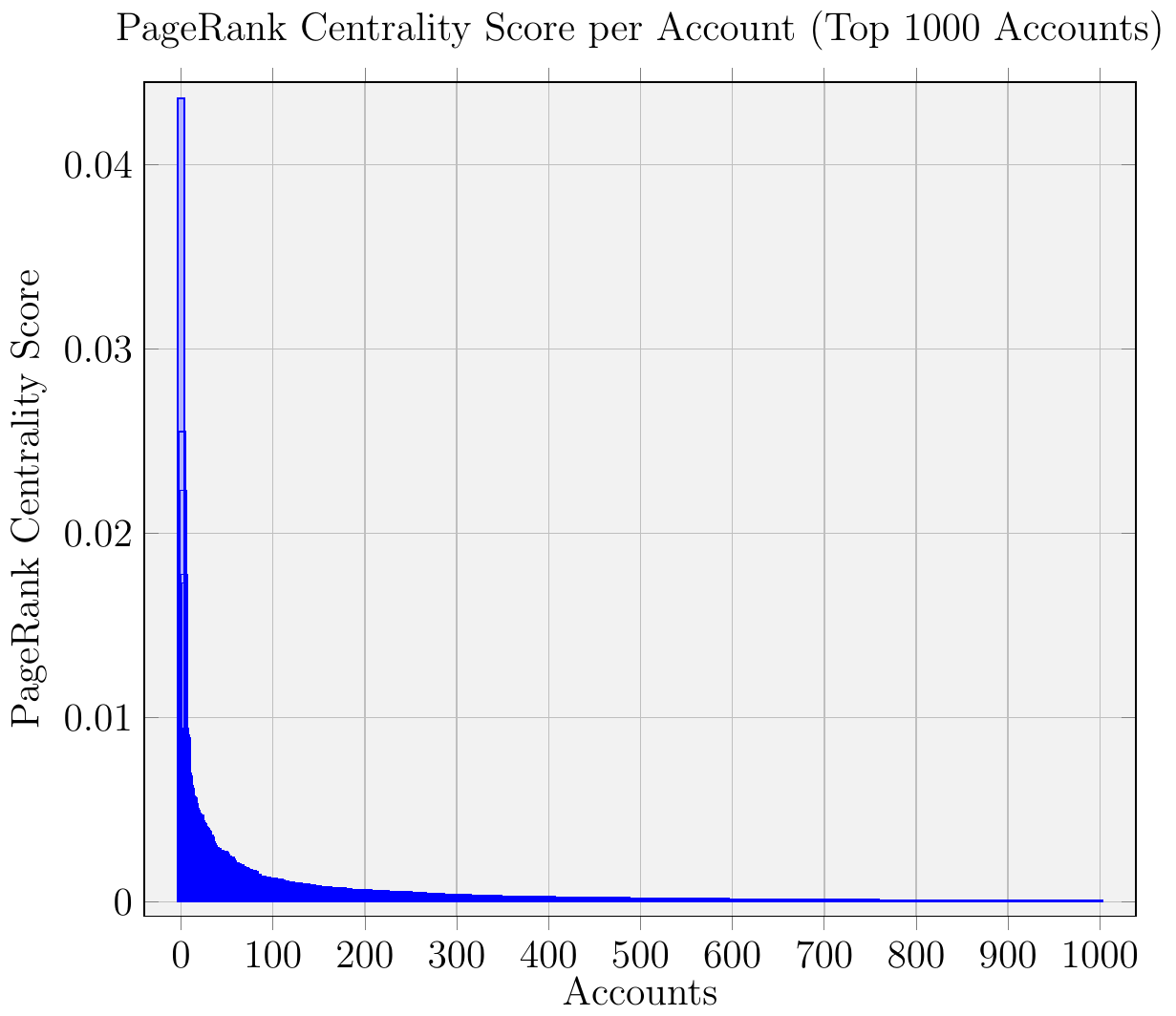}
    \label{fig:pageRank1000}
\end{subfigure}
\quad
\begin{subfigure}[b]{0.4\textwidth}
    \centering
    \includegraphics[width=\textwidth, height=0.21\textheight]{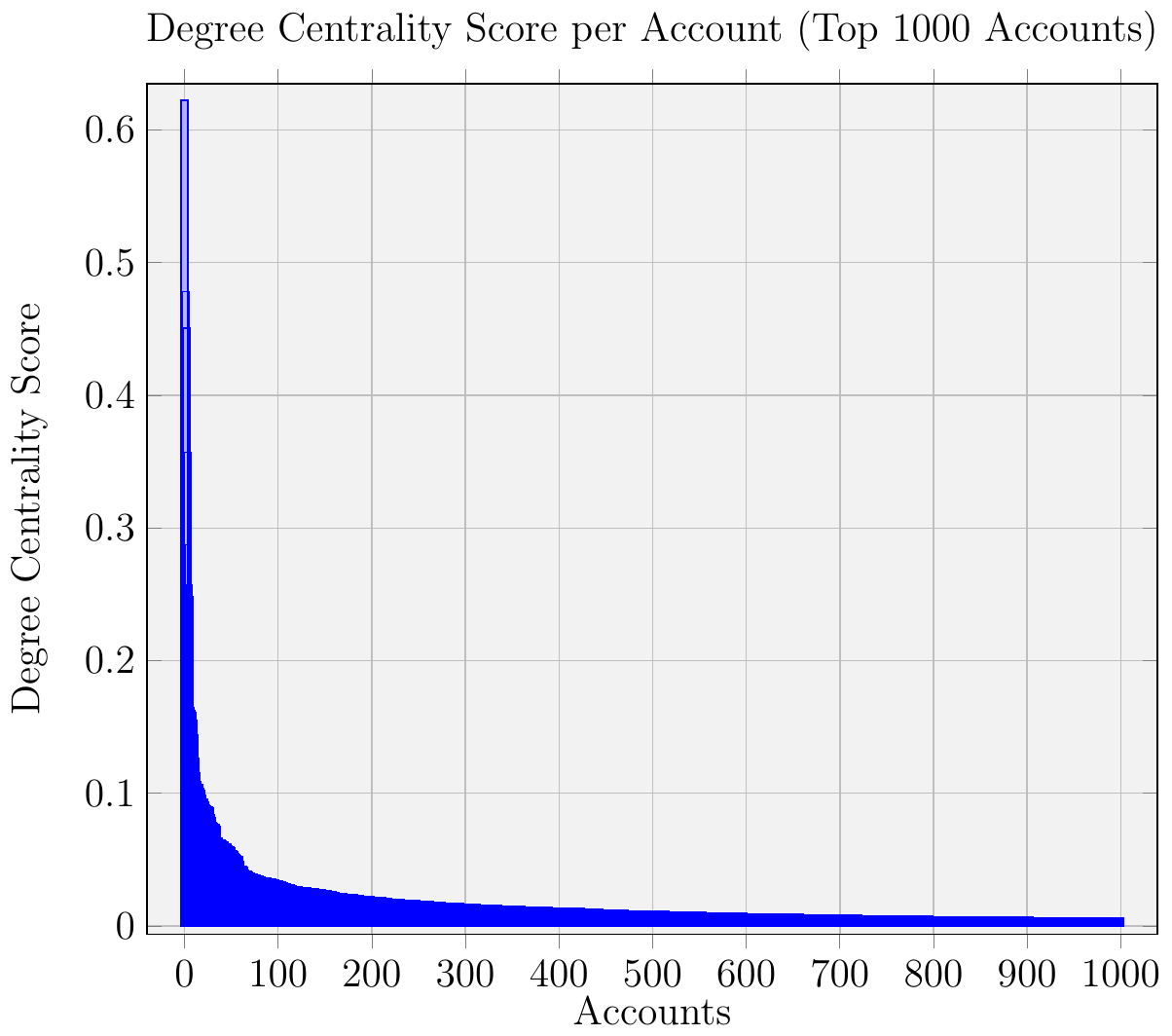}
    \label{fig:degreeCent1000}
\end{subfigure}
\caption{Graphs showing the scores of the 1000 highest scoring accounts in each centrality measure.}
\label{fig:top1000Influencers}
\end{figure*}

After collection, the tweets for each influencer account were pre-processed; including the removal of stop words, expansion of contractions, and the removal of Twitter specific noise like the use of `RT' for retweets. UCI was then used to score topic models with topic numbers starting at 2, and progressing in steps of 6 to a maximum of 40 topics. All coherence scores were again computed using the Python Gensim package~\cite{gensim}.

Upon completion of these calculations, and with reference to their coherence scores, topic numbers 15, 15, 10, 12, 20 and 25 were selected with each topic number linked to each influencer account in descending order of influence respectively (e.g., 15 topics were used for Account A, 15 for Account B, etc.). LDA topic modelling was then carried out on these accounts using the pre-selected topic numbers and the top scoring words for each account identified topic, for each account, were recorded.

\section{Results and Discussion}

In this section, the Anonymous network, and the way in which influence presents itself will be examined and the results of the topic modelling of influencer account tweets will be reviewed and explored in relation to the smaller, qualitative studies of the group. This study will allow us to synthesise the findings of smaller-scale, interview-based studies of Anonymous with our findings of the group on a larger scale. 
We achieved our aims by using SNA and centrality measures to examine how the network presents itself in light of the qualitative studies' claims that the group failed to meet its unifying goal of a leaderless, flat structure~\cite{Uitermark2017,Olson2013}. Furthermore, we consider their further claims that Anonymous has suffered a serious fragmentation and a fall into relative inactivity after the arrest of key members in the period 2011-2013~\cite{Goode2015,Olson2013}. Topic modelling of influencer account tweets is then used to further examine whether the claims by previous qualitative studies that the group is not as ideologically diverse as it appears, manifests itself in the original content of Anonymous affiliates~\cite{Olson2013,Uitermark2017}.

\subsection{Examining Influence in the Anonymous Social Network}

After the two-staged snowball sampling, 20,506 Anonymous accounts, their relevant meta-data (including total number of tweets, number of followers, number of friends, etc.), and their relationships to each other had been identified and collected. This allowed for the creation of a sizeable network graph with 349,445 edges, and in turn not just the identification of influencer accounts, but also an examination of how influence relates to other factors.

In relation to this paper's aim of examining qualitative literature's findings that reject Anonymous' depiction of having no central authority~\cite{Olson2013,Uitermark2017}, this section looks at how influence is distributed across the Anonymous network. This analysis allows us to determine whether a small number of prominent influencers has indeed arisen. These centrality scores are then compared to the creation date of each account to gain a sense of how the network has changed over time, and how influence relates to when an account was created. Moreover, the manner in which the network has evolved over time will be investigated against the findings of smaller-scale, qualitative literature that the group suffered fragmentation after the arrests of prominent members.

\begin{figure}[!b]
\centering
\includegraphics[width=\columnwidth]{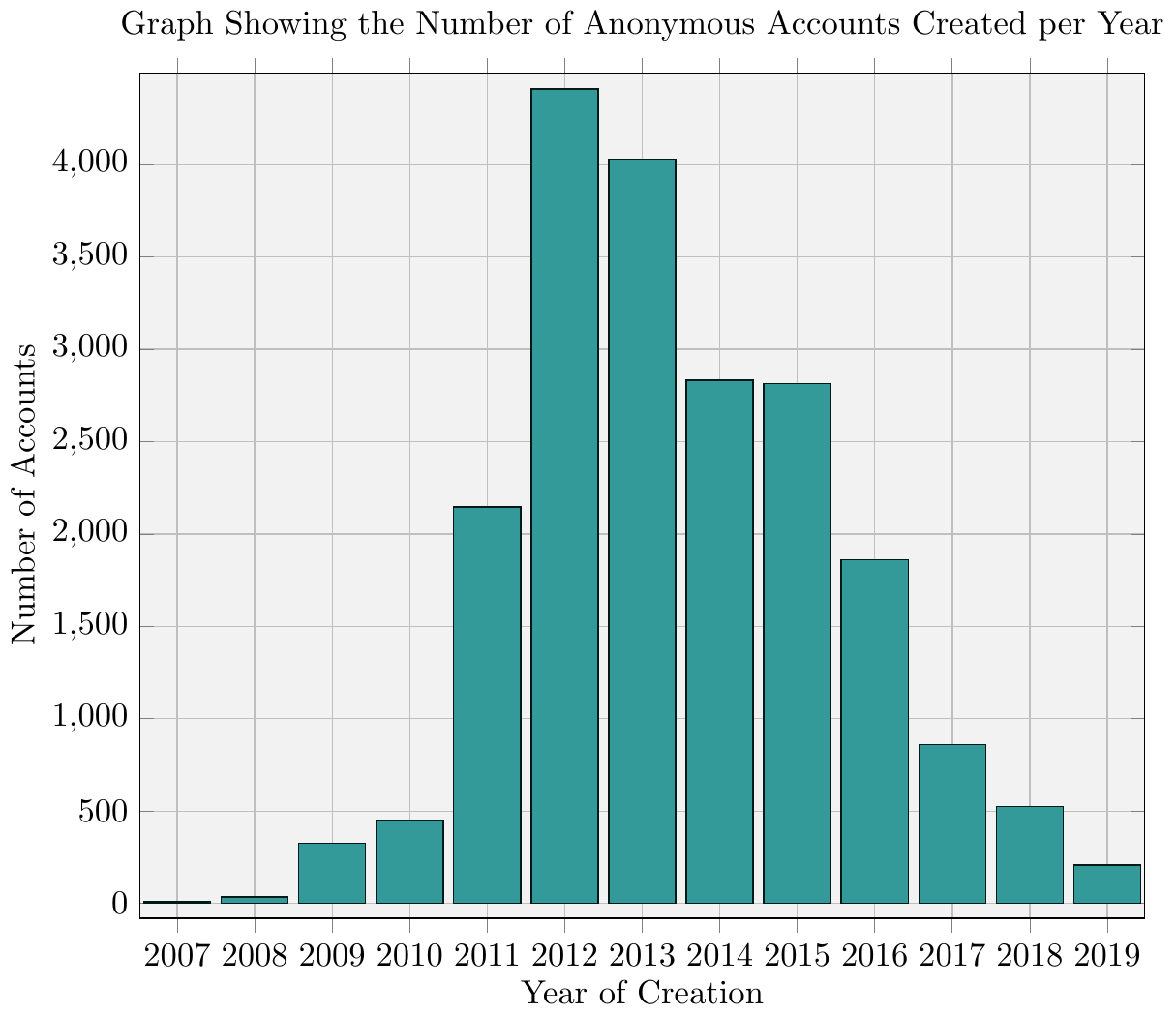}
\caption{Number of Anonymous accounts created each year.}
\label{fig:createDates}
\end{figure}

\subsubsection{Examining the Distribution of Influence}

After the four centrality measures were carried out on this 20,506 strong network, the top scoring accounts from each measure were recorded, as were the top scoring accounts averaged across the four measures. The first area of interest regarding influence was to investigate whether there were a small number of Anonymous accounts scoring highly in influence, or whether influence was more evenly distributed. Across the four metrics, the results were found to be largely similar with a small number of accounts scoring highly for each particular centrality measure, and the rest of the accounts scoring comparatively lowly. A visual depiction of the top 100 scorers for eigenvector and degree centrality can be found in Fig.~\ref{fig:sna_graph}. Moreover, results depicting the top 1,000 accounts for each measure, and their subsequent centrality scores, are defined in Fig.~\ref{fig:top1000Influencers}.

This result is interesting as it appears to reject the claims by the group that they lack any clear hierarchy or set of central figures. Instead, within the Anonymous Twitter network we see a similar balance in terms of influence as the claims made by smaller studies. Ultimately, it appears that the vast majority of influence is actually held by a small number of accounts, whilst the vast majority of Anonymous accounts wield no real influence at all~\cite{Olson2013}. This helps confirm that the notion of a lack of clear structure may be largely theoretical, or perhaps aspirational, within the Anonymous network. Ultimately, an ideal that has not materialised within this newly identified, large network of accounts.

\subsubsection{The Anonymous Network over Time}

To help gain a better understanding of influence within the network, the account creation dates and the last tweet dates were examined for the 20,506 accounts in the network. From this we found that Anonymous accounts displaying the highest centrality scores were generally created between or before 2011-2013 (when the arrests took place), with 62.5\% of the top 500 accounts, and 91.5\% of the top 50 accounts on average having been created in this period. Given the large disparity between the top scoring influencer accounts and the rest of the Anonymous accounts in the network, this indicates that the majority of influencer accounts originated during or before the time of the arrests of the key Anonymous figures. 

There are two things to note, one data-driven and the other contextual, that may help expand upon this finding. Firstly, the accounts present in the Anonymous network were generally created in the years in which the influencer accounts were created (see Fig.~\ref{fig:createDates}). Going backwards in years this is understandable; Anonymous' first real action, ``Operation Chanology'', was carried out in 2008~\cite{Olson2013}. At that time, Twitter was a relatively underused platform, with a mere 6 million users compared to the 117 million it achieved by 2011~\cite{balanceCareers2018}. This increase in users per year matches Anonymous' increased presence on Twitter until peaking around the years 2012 and 2013, when Anonymous account creation slowly appears to drop whilst Twitter's user numbers continued to climb~\cite{balanceCareers2018}.

This continued drop in Anonymous accounts despite the increased popularity of the Twitter platform marries well with the results of the smaller-scale, qualitative studies of the group. These studies' suggestions of a group fragmentation after the 2011-2013 arrests pairs well with the fall in Anonymous accounts being created on Twitter from around the same period of time~\cite{Goode2015,Olson2013}. Moreover, although the popularity of Anonymous on Twitter seems to have ebbed in time with the apparent fragmenting of the group after the arrests, accounts created in Anonymous' period of prominence are still the ones most likely to exert influence. This suggests a more rigid structure of influence than Anonymous has claimed~\cite{Uitermark2017}.

\begin{figure}[!htb]
\centering
\includegraphics[width=\columnwidth]{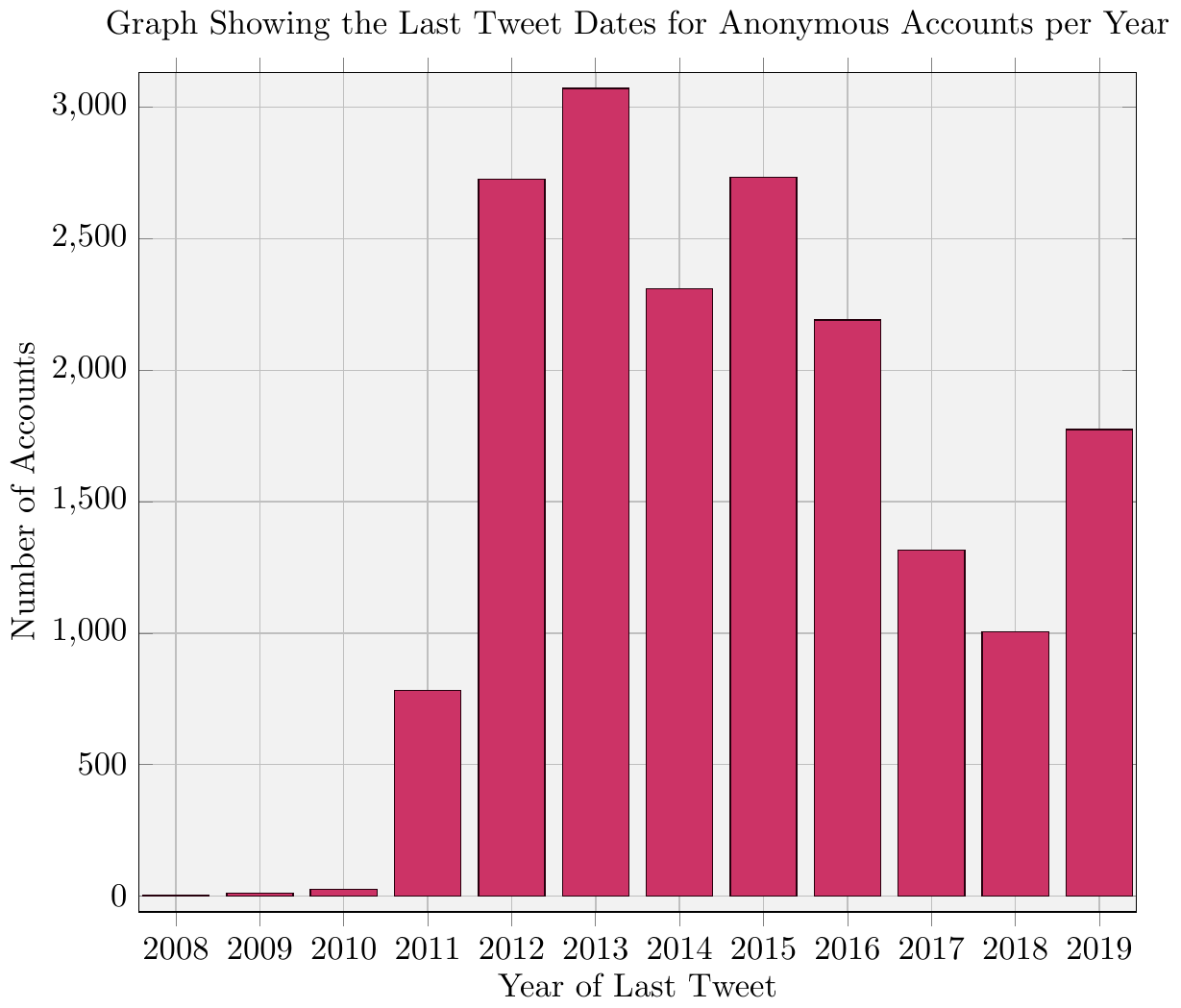}
\caption{Number of Anonymous accounts that stopped tweeting each year.}
\label{fig:lastTweetDate}
\end{figure}

What is worth noting in regards to this more tangible picture moreover, is that of the members of this Anonymous network, only around 50\% have tweeted since 2015, and 35\% since 2016 (Fig.~\ref{fig:lastTweetDate}). This shows that not only has the number of accounts created each year dropped significantly, but also that the majority of accounts have fallen inactive since the arrests. In fact, we can see from Fig.~\ref{fig:lastTweetDate} that only 9\% of accounts have tweeted in the year 2019, and that since 2012 there has been a steady number of accounts tweeting their last tweet. Ultimately, although this network seems to present with a central group of influencers from the pre-arrest period of Anonymous' history, the number of accounts that appeared to be active is but a small fraction of the complete network. The fact that this fall in activity roughly occurs in time with the  arrests in 2011-2013 further indicates that the group suffered a severe fragmentation as a result.

\subsection{Extracting Topics From Anonymous Tweets}

Having extracted the 1,500 latest tweets from the top six scoring Anonymous influencer accounts (four of the top six accounts were accounts previously used as seeds in the account collection phase), topic modelling was carried out to determine the overarching areas of focus present within each account's corpus of tweets. In turn, this allowed us a previously unexamined look into the content of these Anonymous accounts and their overarching topics of focus. This was done in order to examine the cohesiveness and consistency in subject-matter across influencer accounts relative to claims that reject Anonymous' position of having no consistent areas of interest~\cite{Uitermark2017,Olson2013}.

\begin{table*}[!htb]
\centering
\begin{tabular}{p{0.25\textwidth} >{\raggedright\arraybackslash}p{0.65\textwidth}} 
 \toprule
 \textbf{Topics} & \textbf{Key Words} \\ [0.5ex] 
 \midrule
 Conflict in Iraq & people, asylum, concern, iraq, issue, turn, hate, learn, internet, question \\
 \midrule
 Donald Trump and Conflict & trump, war, crime, free, people, war\_crime, time, election, feel, platform \\
 \midrule
  Democrats and US Politics & democrat, post, expect, voter, exit, profit, server, goal, poll, send \\
 \midrule
 Republicans and US Politics & problem, person, republican, american, identity, challenge, obama, understand, steal, order \\
 \midrule
  US Politics & administration, prosecute, america, ago, pass, woman, article, long, good, bush \\
 \midrule
  Brexit & expose, ireland, border, country, good, remain, life, journalist, return, password \\
 \midrule
  Journalistic Freedom & freedom, press, press\_freedom, people, chelseamanning, thing, bit, claim, fascist, white \\
 \midrule
  EU Copyright Directive & vote, eucopyrightdirective, european, medium, democracy, response, war, uploadfilters, happen, article13 \\
 \midrule
 Wikileaks and Assange's Arrest & arrest, ecuador, people, human, wikileaks, wrong, publish, time, block, hold \\
 \midrule
 Wikileaks and Edward Snowden & assange, torture, government, charge, wikileaks, read, whistleblower, julian, snowden, julian\_assange \\
 \midrule
  Anonymous and Anon Members & [\textit{Redacted}], contributor, [\textit{Redacted}], [\textit{Redacted}], account, content, fail, [\textit{Redacted}], [\textit{Redacted}] \\
  \midrule
 Twitter & account, twitter, live, publish, require, day, long, user, forget, simple \\
 \midrule
 \textit{Unclear} & brexit, people, fact, manning, year, file, power, pay, party, work \\
  \midrule
 \textit{Unclear} & history, work, mind, privacy, laugh, flee, data, document, politics, lose \\
  \midrule
 \textit{Unclear} & tweet, leak, link, access, state, change, expose, political, bad, anti \\
 \bottomrule
\end{tabular}
\caption{Account A's Topics. Certain keywords have been redacted as they identified active Twitter accounts.}
\label{table:accountATopics}
\end{table*}

\begin{figure}[!htb]
\centering
\includegraphics[width=\columnwidth]{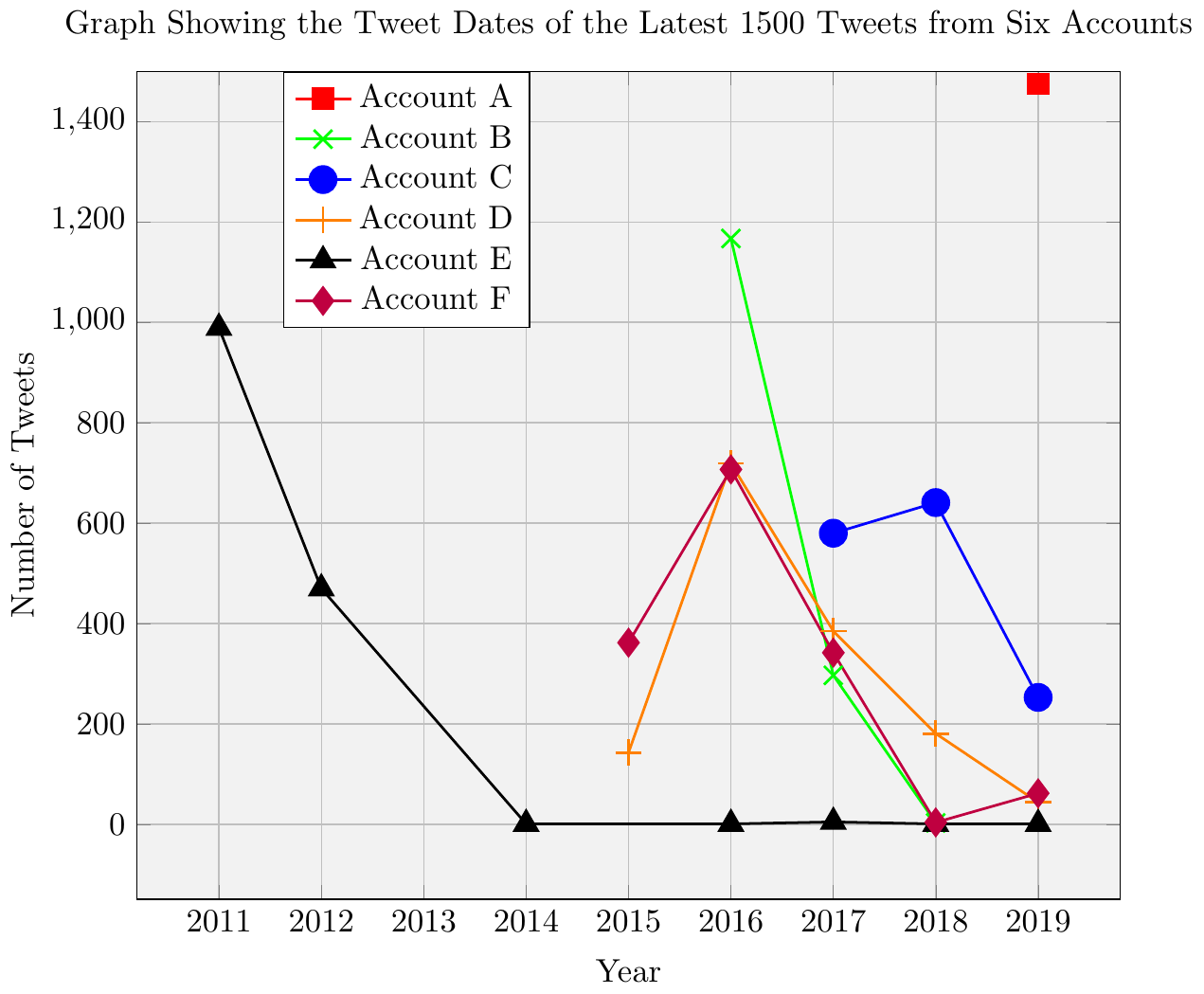}
\caption{Time series graph showing each influencer account, and the years in which their 1,500 most recent tweets were tweeted.}
\label{fig:timeSeriesTweet}
\end{figure}

Before continuing further into the results of the topic modelling, it is first worth noting that due to the differing tweeting habits of each of these six accounts, the period of time captured within each batch of 1,500 tweets varies between accounts (see Fig.~\ref{fig:timeSeriesTweet}). 
It also worth restating the number of topics with which each account's corpus of tweets was divided. Account A used 15 topics, Account B: 15, Account C: 10, Account D: 12, and Account E: 20, and Account F: 25.

After topic modelling was carried out, each topic was labelled in accordance with the top ten words belonging to that topic, and the likely theme, person, event, etc.\ it represented. An example of this can be seen in Table~\ref{table:accountATopics}, showing the topics and key words found from Account A's tweets.

\subsubsection{Topic Modelling Results}

For Account A, the majority of topics are focused on politics in some fashion, generally centred on the politics of the USA and the UK. This includes general topics about American politics, Brexit, and Donald Trump. There is also one topic concerned with the recent EU Copyright Directive, whose Articles 11 and 13 have brought recent controversy regarding the requirement for online platforms ``to filter or remove copyrighted material from their websites''~\cite{wired2019}. Account A was also the only account to contain topics that included within its key words the Twitter names of known associates of Anonymous (these names have been anonymised).

Account B's topics also contained topics focused both generally, and specifically on US politics, alongside topics focused on net neutrality. Although we cannot say for certain given the nature of this analytical method, bearing the account's time span in terms of collected tweets (2017-2019) in mind, this likely is in reference to the repeal of net neutrality that occurred in 2018. Indeed, the appearance of `FCC' as a keyword for one of the net neutrality topics strengthens this assumption. The FCC (Federal Communications Commission) were responsible for voting for the repealing of net neutrality policies in December 2017 -- a period of time covered by the tweets collected from this account~\cite{Verge2019}. There were also two separate topics likely focused on significant data breaches affecting Facebook and Equifax~\cite{Guardian2018,BBCNews2019}.

Account C's topics contained references to US politics, in particular the 2016 general election (this account's tweets cover that time span). However, the majority of topics are focused either on Wikileaks, leak culture in general, or political action. Of interest is the presence of three separate identified topics which seem focused on Wikileaks and Turkey; presumably -- given the appearance of the words `coup' and `erdogan' -- the publishing by Wikileaks of the `Erdogan Emails' directly after the failed coup in 2016~\cite{wired2016}. Also of note is the presence of the topic concerning `Op Icarus', the only Op related topic identified within these accounts' tweets. Op Icarus was an action carried out by Anonymous, and involved the use of DDoS attacks against eight international banks' websites in May 2016~\cite{IBT2016}.

Account D's topics presented in a similar way, with the majority focused on US politics --  Donald Trump in particular -- as well as topics indicating discussion of political change; and wider topics focused on leaks, data security, backdoors and encryption. Of note is a topic focused around Edward Snowden -- the former NSA analyst responsible for leaking top-secret NSA documents to The Guardian newspaper -- and the words `paypal' and `donation'. Again, one has to be careful in making inferences from this very high level view of tweets containing each topic, but it is of note that `Edward Snowden's website actively seeks donations to `Help cover the defence costs of the whistle-blower who revealed the NSA mass surveillance programs', and that these donations can be sent through Paypal~\cite{Snowden2020}.

Account E's topics also share a political focus, and as well as discussions based around activism, protest, and the Occupy movement. Specifically, the emphasis falls on police brutality, as well as protests against Acta, an intellectual property rights agreement that some feared would lead to censorship on the web~\cite{Wired2012}. In line with the other accounts, there was also topics focused on leaking; including discussion of Julian Assange and Wikileaks. This includes reference to geopolitical intelligence company Stratfor, and presumably the leaking of their emails to Wikileaks~\cite{Insider2012}.

Finally, Account F once again shares great similarities with the other accounts. Topics from this account include a focus on politics in general, as well as specifically on Brexit, Donald Trump, and Turkey -- particularly the 2016 coup. Topics referencing leaking also appear, with reference to Wikileaks and Edward Snowden. 

As a whole, we can see a number of consistencies appearing across each of the Anonymous influencer accounts. It appears that all of the accounts share a focus on political matters, especially those of the USA, with a keen focus on President Trump. Moreover, issues concerning data are also shared by all of the accounts, and issues concerning the leaking of data appear in all but one (Account B). In turn, we also have to remember that due to the differing time frames, the lack of a particular topic from one account to the next does not necessitate that that topic was not discussed. Issues such as net neutrality are largely focused around the years 2017 and 2018~\cite{Verge2019}, meaning that although it only appeared in one of the account's models, other accounts, like Account A, may have still discussed it. It is just that the time frame for their latest 1,500 tweets (the year 2019 for Account A), rendered this topic less likely to occur. 

Therefore, reflecting on the cohesiveness in influencer account content, we certainly see within these topics Anonymous influencers largely centred on similar subject matter both to each other, and to the `active' Anonymous of the early 2010s~\cite{Olson2013,Uitermark2017}. Although in theory Anonymous claims to be a platform to be used by any, for anything; for one reason or another this does not seem to be the case on Twitter, at least for these accounts of influence. The vast majority of topics identified are largely in line with the forms of action the group (particularly the `moralfag' arm of the group) was taking during its more notorious periods in the early 2010s, and are largely in line with each other~\cite{Olson2013}. Their focus is largely extended to politics, political action, the freedom of information, leaks, cyber-security, and very little else. Issues that seem to fit nicely with the content one would expect from Anonymous affiliates during the prime years of the group~\cite{Olson2013,Uitermark2017}.

Moreover, as some of these collections of tweets are spread throughout several years, we are offered a more long-term sense of these accounts' activity. From this, it seems that accounts tweeting over a long period of time are tweeting about roughly the same subject matter as those tweeting over a shorter period. This finding indicates that these consistencies present in both the long-term tweeting habits of these influencer accounts, as well as the short -- further contradicting the claims of the group to its ephemeral nature~\cite{Uitermark2017}. 

Furthermore, a feature that was surprisingly absent, but which unified the topics of these accounts, is the lack of `lulz' topics -- one of the key ideological rifts noted in the group's more notorious period before the 2011-2013 arrests~\cite{Uitermark2017}. This lack indicates a unity between these accounts and potentially the network as a whole, as well as a notable absence of this facet of Anonymous. Whilst the `lulz' may live on in other networks, or other parts of this network, it is worth highlighting that the key influencer accounts are united towards topics of a more high-minded value. This therefore indicates a distancing, in this particular network, from the more nihilistic aspects often associated with the group. 

Also surprising is the apparent lack of Op related content. Although some of these topics may have some form of Op attached to them, only one topic -- Op Icarus from Account C -- has an Op present in its keywords. It is possible that the lack of Op mentions mirrors the suggested inactivity and fragmentation of Anonymous post-arrests.

\subsection{Limitations}

There are a few limitations with regards to the data used that need to be acknowledged. In terms of the Anonymous network, it may be worth expanding upon a two-staged snowball sample to examine the full extent of the network. More work could thus be done to expand upon this work to identify the network in its entirety, or indeed to identify the presence of other Anonymous networks that may exist separately from the one found in this paper. Moreover, due to certain accounts (e.g., those suspended, deleted or protected) being made unavailable, we would like to acknowledge that whilst our study examines a large number of Anonymous accounts, it does not necessarily offer a complete picture of the network in its entirety. This is, however, more a reflection of the limitations of Twitter's Standard search API, rather than the project itself. 

In regards to the topic modelling, a larger scale project over a greater period of time would allow for a higher number of tweets collected per account, as well as a higher number of influencer accounts included in the topic modelling. This would help improve our understanding of the similarities and differences in content between influencer accounts, and strengthen the conclusions made by this paper.

\section{Conclusions}

Overall, our large-scale study of the Anonymous network on Twitter has worked to provide a synthesis with the findings of smaller qualitative studies of the Anonymous collective. In turn providing their findings, largely derived from interviews, with evidence from the large scale behaviour of self-identified Anonymous affiliates.

In our studies of influence, we have found that whilst Anonymous makes claims to a flat structure, this may not be the case. Just as the smaller-scale studies suggested, the reality for this large Anonymous network is that influence appears to be the privilege of a very small number of accounts. Accounts created during Anonymous' period of more frequent operation before the arrests of their `key' members~\cite{Uitermark2017,Olson2013}. Moreover, we see that the number of Anonymous accounts being created each year has decreased significantly, and the number of accounts falling silently each year is significant too. 

Furthermore given that the key influencers on the Anonymous network are accounts created during the group's more prominent period, it is not necessarily surprising that the content we see from them is very much in line with what the group was concerned with during the early 2010s -- particularly focusing on political action and the sharing of information. In addition, the surprising lack of Op based tweets potentially supports the fragmentation and fall to inactivity suggested by qualitative studies of the group~\cite{Uitermark2017,Olson2013,Goode2015}.

\subsection{Future Work}

In future, work could be done to expand the scope of the research beyond six influencer accounts to look at the content of the Anonymous network as something more akin to a whole. This would help us understand whether the similarity in content of the influencer accounts is something that appears across the network in its entirety. Moreover, additional metrics of influence may be useful to verify the conclusions drawn from the centrality measures. Finally, a larger study of Anonymous account topics over a longer period could help to verify whether the lack of Op based content indicates a fall in the group's activity, or perhaps that the Anonymous Twitter network has never involved itself in this aspect of the group.

Further work could also be carried out to compare the findings of this paper with the behaviours of Anonymous on other social media platforms, such as Facebook and YouTube, as well as with their more private dealings on IRC. These studies could help us gain a broader sense of the group's behaviours, and whether the findings of this paper are purely a description of Anonymous on Twitter, or whether they are also emergent in the group's general behaviours on other platforms.

Finally, the methods detailed here -- combining machine learning for identification of affiliates, with SNA and topic modelling to examine behaviour -- could also be repurposed towards the study of other groups online. Due to the relative simplicity and intuitive nature of the methods, these could even be of potential use to law enforcement and the cyber threat intelligence community who are interested in gaining a broader understanding of criminal organisations online.


\selectfont
\bibliographystyle{aaai}
\bibliography{FULL-JonesK} 

\end{document}